\documentclass[11pt, letter]{article}

\usepackage{stolenstyle}

\usepackage{amsmath,epsf,amssymb,latexsym,amsthm,setspace,bbm,array,pifont,enumerate}

\usepackage[matrix,arrow,frame,import,curve,color]{xy}

\DeclareFontFamily{U}{rsf}{}
\DeclareFontShape{U}{rsf}{m}{n}{
  <5> <6> rsfs5 <7> <8> <9> rsfs7 <10-> rsfs10}{}
\DeclareMathAlphabet\Scr{U}{rsf}{m}{n}

\def\cDb{{\overline{\cD}}}

\def\GUL{\GU(1)_{\text{L}}}
\def\GUR{\GU(1)_{\text{R}}}

\def\C{{\mathbb C}}

\def\R{{\mathbb R}}
\def\Z{{\mathbb Z}}

\def\SU{\operatorname{SU}}
\def\GU{\operatorname{U{}}}

\def\GE{\operatorname{E}}

\def\Lsl{\operatorname{\mathfrak{sl}}}

\def\p{\partial}
\def\pb{\bar{\partial}}

\def\la{\langle}
\def\ra{\rangle}

\def\ff#1#2{{\textstyle\frac{#1}{#2}}}

\def\cD{{\cal D}}

\def\cL{{\cal L}}
\def\cM{{\cal M}}

\def\cW{{\cal W}}

\def\cZ{{\cal Z}}

\newcommand\gammab{\overline{\gamma}}

\newcommand\thetab{\overline{\theta}}

\newcommand\phib{\overline{\phi}}

\newcommand\psib{\overline{\psi}}

\newcommand\Gammab{\overline{\Gamma}}

\newcommand\Phib{\overline{\Phi}}

\newcommand\cb{\overline{c}}

\newcommand\hb{\overline{h}}

\newcommand\qb{\overline{q}}

\newcommand\zb{\overline{z}}

\theoremstyle{definition}

\usepackage{tikz}

\usetikzlibrary[shapes.geometric]
\usetikzlibrary{positioning} 
\usetikzlibrary{calc,intersections,through,backgrounds}

\tikzset{>=stealth}
\tikzset{every picture/.style={very thick}}

\def\cWb{{\overline{\cW}}}

\def\bQb{{{{\overline{\boldsymbol{Q}}}}}}

\def\preprint{ ~~ }

\title{Small Landau--Ginzburg theories}
\author {Sean M.~Gholson and Ilarion V.~Melnikov}
\affiliation{Department of Physics and Astronomy,\\
James Madison University, Harrisonburg, VA 22807, USA}
\emailAdd{gholsosm@dukes.jmu.edu, melnikix@jmu.edu}

\abstract{We classify (0,2) Landau-Ginzburg theories that can flow to compact IR fixed points with equal left and right central charges strictly bounded by $3$.  Our result is a (0,2) generalization of the ADE classification of (2,2) Landau-Ginzburg theories that flow to N=2 minimal models.  Unitarity requires the right-moving supersymmetric sector to fall into the standard N=2 minimal model representations, but the left-moving sector need not have supersymmetry.  The Landau-Ginzburg realizations provide a simple way to compute the chiral algebra and other characteristics of these fixed points.  While our results pertain to isolated superconformal theories, tensor products lead to (0,2) superconformal theories with higher central charge, and the Landau-Ginzburg realization provides a model for a class of marginal and relevant deformations of such theories.}

\begin{document}

\maketitle

\section{Introduction}\label{s:intro}

The super Virasoro minimal models are gems in the vast landscape of two-dimensional superconformal field theories (SCFTs).  They are exactly solvable, and, while comparatively simple, their tensor products and orbifolds can be used to build rich families of superconformal theories with exactly marginal deformations and a corresponding moduli space of theories.  In the case of left-right symmetric theories without supersymmetry or with (1,1) superconformal invariance, the minimal models arise as IR fixed points of RG flows from Landau-Ginzburg (LG) Lagrangian theories of fermions and scalars with interactions determined by a superpotential~\cite{Zamolodchikov:1986db,Kastor:1988ef}.    

In the (2,2) case, there is a beautiful correspondence between ADE quasi-homogeneous singularities~\cite{MR2896292} and the (2,2) minimal models~\cite{Boucher:1986bh} described in~\cite{Martinec:1988zu,Vafa:1988uu}.  The weakly coupled UV description is perfectly suited to computing a number of RG--invariant quantities, which reflect properties of the superconformal theory.  In addition, the Lagrangian presentation is well-suited to describing tensor products of minimal models.  For instance, certain marginal deformations of the superconformal theory are encoded by the holomorphic parameters of the Lagrangian, which allows for an exact determination of the dependence of protected correlation functions of the SCFT~\cite{Vafa:1990mu}.

Taking a somewhat different perspective, one can ask: ``which superconformal field theories arise as IR fixed points of LG theories?''  This is a fruitful question because the simplicity of the LG description is not restricted to the minimal models:  a superconformal theory with a LG description can be studied by straightforward generalizations of the technology applicable to the minimal models.  The answer depends on what one means by a LG theory.  We will use the classical meaning of the  term:  a Lagrangian theory with a free kinetic term for fermions and bosons and a scalar potential with isolated critical points (we will call such a potential ``compact'').  In the (2,2) superconformal case we will also insist that the superpotential is a polynomial and quasi-homogeneous function of the chiral superfields, so that the Lagrangian theory has a manifest chiral R-symmetry, and we will assume that the current for this R-symmetry flows to the current for the R-symmetry of the IR SCFT.  With these assumptions, it is well-known that for (2,2) theories there is a finite number of LG theories that realize any fixed rational central charge $c=\cb$~\cite{Kreuzer:1992bi}, and it is practical to enumerate the theories when $c$ is not too large.  For instance, the classification of (2,2) LG theories that flow to IR theories with $c=\cb=9$ (which played an important role in the early days of mirror symmetry and the Calabi-Yau/LG correspondence of stringy geometry) yields 10,839 theories distinguished by the spectrum of chiral primary operators~\cite{Kreuzer:1992np,Klemm:1992bx}.  Using the methods developed in those classifications, it is also possible to enumerate the theories with central charge smaller than some fixed bound.  For instance, for $c<6$ such an enumeration and its relation to the theory of quasi-homogeneous singularities is given in~\cite{Davenport:2016ggc}.

The (2,2) LG theories belong to a wider class of models that only preserve (0,2) worldsheet supersymmetry.  Indeed, the (2,2) Lagrangian theories typically have continuous deformations that preserve (0,2) supersymmetry and the chiral R-symmetry, and thus can describe (0,2)--preserving deformations of (2,2) SCFTs.  More generally, it has long been known that (0,2) LG theories act as useful signposts in the landscape of (0,2) SCFTs~\cite{Witten:1993yc,Distler:1993mk,Kawai:1993jk,Blumenhagen:1995ew}.  Despite this promise and many applications in particular examples, the (0,2) LG theories are less understood than their (2,2) relatives.  For instance, it is not known how to enumerate (0,2) LG theories that realize some fixed central charges $c,\cb$, nor has it been shown that the number of such theories is finite.  The classification is complicated by accidental symmetries that can invalidate the naive correspondence between the UV R-symmetry and the R-symmetry of the IR SCFT~\cite{Bertolini:2014ela}.

In this work we present the classification of (0,2) LG theories with $c=\cb<3$.  We find two infinite families and three sporadic (0,2) LG theories with $c=\cb<3$; the results are listed in~(\ref{eq:thefinalcut}).  This is a very natural class of theories to examine.  First, the strong unitarity constraints of the $\cb < 3$ N=2 superconformal algebra offer a hope that the classification problem will be tractable.  Second, the IR fixed points obtained in this fashion help to fill out the space of ``(0,2) minimal models''---a notion we will discuss further below.  Indeed, some of our families have been identified with exactly solvable (0,2) SCFTs~\cite{Gadde:2016khg}.  Finally, just as for (2,2) theories and their deformations, by taking tensor products of the theories we can build many (0,2) SCFTs that are amenable to study through the LG UV description.  For instance, these can act as new ingredients in constructing (0,2) SCFTs suitable for heterotic compactification.

The rest of this note is organized as follows.  In section~\ref{s:overview} we review some (0,2) LG and minimal model notions relevant to our study.  In section~\ref{s:justtwo}, we show that unitarity constrains the analysis to theories with just one or two fermi multiplets, and in the following section we enumerate the families that are consistent with our charge bounds and compactness.  Next, in section~\ref{s:classredef}, we study field redefinitions to prove a number of isomorphisms that allow us to enumerate the inequivalent theories.  We conclude with a short discussion of the results.

\section*{Acknowledgements}  IVM would like to thank M. Bertolini for useful discussions and M.R.~Plesser for comments on and corrections to the manuscript. The work of SMG was supported in part through the Tickle Foundation, and the work of IVM was supported in part through the 4-VA grant ``Frontiers in stringy geometry.''

\section{(0,2) Landau-Ginzburg overview} \label{s:overview}

\subsection{(2,2) LG and minimal models}
To set the stage for our work, we begin with a quick overview of (2,2) LG theories with $c=\cb<3$~\cite{Kreuzer:1992bi}.  The Lagrangian has $n$ chiral multiplets $X_i$ with free kinetic terms and a quasi-homogeneous (2,2) superpotential $W(X_1,\ldots, X_n)$ satisfying
\begin{align}
W(t^{q_1} X_1, t^{q_2} X_2, \ldots, t^{q_3} X_n) = t W(X_1,\ldots, X_n)~.
\end{align}
The $q_i$ are rational R-charges satisfying $0< q_i < 1/2$.  The lower bound follows from unitarity and compactness requirements on the SCFT: each $X_i$ should flow to an operator with R-charge $q_i$, and we focus on SCFTs with a normalizable $\Lsl(2,\C)$ vacuum.  The upper bound follows because fields with $q_i = 1/2$ correspond to massive multiplets:  these can be integrated out without affecting the IR fixed point.  

Once the $q_i$ are fixed, and we make the assumption that the UV R-current flows to the IR R-current, the central charge can be determined in a variety of ways.  For instance, one might study the Weyl rescaling of the world-sheet metric~\cite{Vafa:1988uu}, the elliptic genus or the chiral algebra~\cite{Lerche:1989uy,Witten:1993jg},  or use c-extremization~\cite{Benini:2012cz}.  The result is the famous expression
\begin{align}
c=\cb = 3 \sum_{i=1}^n (1- 2 q_i)~.
\end{align}
The critical locus of the superpotential is the set of solutions to
\begin{align}
\frac{\p W}{\p X_1} & = 0~,&
\frac{\p W}{\p X_2} & = 0~,& \ldots~, &&
\frac{\p W}{\p X_n} & = 0~.
\end{align}
Since $W$ is quasi-homogeneous, compactness implies that the critical locus consists of the origin $X_1 = X_2 = \ldots =X_n=0$, and with some work it can be shown that this requires $c \ge n$~\cite{Kreuzer:1992bi}.

At this point, the correspondence with the (2,2) minimal models~\cite{Boucher:1986bh} is quite natural, because it is easy to describe the compact quasi-homogenous superpotentials with $n \le 2$---a classic result in singularity theory.  Since we will be doing similar manipulations below, we will take a moment to review the argument.

First we consider the case of $n=1$.  In this case the only quasi-homogenous potential is $W = X_1^{k}$, which yields $c = 3 (k-2)/k$.  Next, we let $n=2$.  It is an amusing exercise to show that a compact superpotential necessarily takes one of the following three forms:
\begin{align}
W_1 & \supset X_1^{k_1} + X_2^{k_2} & \implies  q_1 & = \frac{1}{k_1}~,&  q_2 & = \frac{1}{k_2}~,& c_1 & = 6 \left(1-\frac{1}{k_1}-\frac{1}{k_2}\right)~,\nonumber\\
W_2 & \supset X_1^{k_1} + X_1 X_2^{k_2} & \implies q_1 & = \frac{1}{k_1}~,& q_2 &= \frac{k_1-1}{k_1k_2}~,&  c_2 & = 6\left(1-\frac{1}{k_1}\right)\left(1-\frac{1}{k_2}\right)~, \nonumber\\
W_3 & \supset X_1 X_2^{l_2} + X_1^{l_1} X_2 & \implies q_1 & = \frac{l_2-1}{l_1l_2-1}~,&q_2 &= \frac{l_1-1}{l_1l_2-1}~,& c_3 & = 6 \left(1-\frac{l_1+l_2-2}{l_1l_2-1}\right)~.
\end{align}
More precisely, unless the shown monomials appear with generic coefficients, the critical locus will be non-compact; on the other hand, once the indicated monomials are present, they uniquely determine the charges $q_1$, $q_2$, and therefore $c$.

We have yet to impose the requirement $c <3$ obeyed by the N=2 unitary minimal models.  For the first class of superpotentials we need
\begin{align}
\frac{1}{k_1} + \frac{1}{k_2} >\frac{1}{2}~,
\end{align}
and the solutions are, together with the ADE labels,
\begin{align}
\label{eq:AAA}
\text{A}_1 \oplus \text{A}_1 && W &= X_1^3 + X_2^3 &\frac{c}{3} &= \frac{4}{4+2}~,\nonumber\\
\text{A}_1 \oplus \text{A}_2 = \text{E}_6 && W &= X_1^3+X_2^4 & \frac{c}{3} &= \frac{10}{10+2}~,\nonumber\\
\text{A}_1 \oplus \text{A}_3 = \text{E}_8 && W &= X_1^3 + X_2^5 & \frac{c}{3} &= \frac{28}{28+2}~~.
\end{align}
We wrote the central charges in the slightly strange form to emphasize that they take the form $c=\frac{3k}{k+2}$ for integer $k$, as is appropriate for the unitary minimal models.  Note that we did not impose this by hand on the LG charges:  merely requiring $c<3$ forces the possible solutions to satisfy the basic unitarity constraint.

Proceeding to the remaining classes of superpotentials, we find that for $W_2$ the constraint $c<3$ leads to 
\begin{align}
\label{eq:DE}
\text{D}_{k+1} && W & = X_1^{k_1} + X_1 X_2^{2} &\frac{c}{3} & = \frac{ 2k_1-2}{2k_1-2+2}~,\nonumber\\
\text{E}_7 && W & = X_1^3 + X_1 X_2^3 &\frac{c}{3} & = \frac{16}{16+2}~.
\end{align}
For $W_3$ we can, without loss of generality, take $l_1\ge l_2$, and then $c<3$ is satisfied by
\begin{align}
W & = X_1 X_2^2 + X_1^{l_1} X_2~.
\end{align}
If the picture of the RG flow just presented is to be sensible, this class should also correspond to some minimal model.  That is indeed the case, and it can be seen explicitly through a holomorphic field redefinition compatible with the R-charges:
\begin{align}
X_1 & = Y_1~ ,\nonumber\\
X_2 & = Y_2 + \alpha Y_1^{l_1-1}~,
\end{align}
which leads to
\begin{align}
W & = -\frac{1}{4} Y_1^{2l_1-1} + Y_1 Y_2^2~,
\end{align}
the superpotential for a $D_{2l_1}$ minimal model.  The Lagrangians for the LG theories with $W_3$ and $c<3$ differ from those with $W_2$ and $c<3$ by D-terms; we assume these to be irrelevant, so that if our interest is in the distinct IR fixed points obtained from LG theories, then the superpotentials in ~(\ref{eq:AAA},\ref{eq:DE}) describe all of the possibilities.  Finally, it also not hard to show that these superpotentials are rigid:  any continuous parameter in $W$ compatible with the R-charges can be absorbed into a holomorphic field redefinition.

The same logic will allow us to classify the $c=\cb <3 $ (0,2) LG theories up to field redefinitions.  However, before we launch into those details, we will review our (0,2) conventions, following~\cite{Bertolini:2014ela}.

\subsection{(0,2) LG conventions}
A (0,2) LG theory is a Lagrangian theory with two types of (0,2) supermultiplets:  the chiral bosonic multiplets $\Phi_i$, $i=1,\ldots, n$, each containing a complex boson $\phi_i$ and a right-moving fermion $\psi_i$, and the chiral fermi multiplets $\Gamma_A$, $A=1,\ldots, N$, each containing a left-moving fermion $\gamma_A$ and a complex auxiliary field.  For each chiral multiplet the theory also includes the antichiral conjugate multiplet.

 The Lagrangian is conveniently presented in terms of (0,2) superspace with superspace coordinates $(z,\zb;\theta,\thetab)$ and superspace derivatives $\cD = \p_{\theta} +\thetab \pb$, $\cDb= \p_{\thetab} + \theta\p$.  It is a sum of two terms
\begin{align}
\cL = \cL_{\text{kin}} + \cL_{W}~,
\end{align}
with
\begin{align}
\cL_{\text{kin}} &= \left. \cD \cDb \left[ \ff{1}{2} \Phib_i \p \Phi_i - \ff{1}{2} \Gammab_A \Gamma_A\right]\right|_{\theta,\thetab=0}~, &
\cL_{W} & =  \left. \left[\cD \cW + \cDb \cWb\right]\right|_{\theta=\thetab=0}~,  \nonumber\\
\cW & = \sum_{A} \Gamma_A J_A(\Phi)~.
\end{align}
Our main interest is in the holomorphic (0,2) superpotential $\cW$, which is determined by $N$ holomorphic polynomials $J_A(\Phi)$.  The (0,2) scalar potential is proportional to $\sum_A \|J_A\|^2$ and will be compact if and only if the the common vanishing locus of the $J_A$ consists of isolated points.  In that case we say the corresponding ideal $\mathbb{J} = \la J_1,\ldots, J_N\ra \subset \C[\phi_1,\ldots,\phi_n]$ is zero-dimensional.   We will be interested in theories that preserve a $\GUL\times\GUR$ symmetry\footnote{In principle we need only require a $\GUR$ symmetry; however, in this class of theories a $\GUL\times\GUR$ UV symmetry is necessary to construct a candidate IR R-symmetry consistent with $\cb >0$.}, of which the former is a global symmetry of the superpotential $\cW$:
\begin{align}
\GUL:&&  \Gamma_A &\to e^{iQ^A \lambda} \Gamma_A  & \Phi_i &\to e^{i q_i\lambda} \Phi_i~, 
\end{align}
which requires $J_A$ to be quasi-homogeneous of degree $-Q_A$:
\begin{align}
J_A(e^{i q_i \lambda} \Phi_i) = e^{-iQ_A \lambda} J_A(\Phi)~.
\end{align}
We will assume that the superpotential couplings are generic enough so that there is a single linearly-independent quasi-homogeneous relation:  the $\GUL$ charges are determined up to an overall scaling.  The $\GUR$ symmetry assigns charge $+1$ to the superspace coordinate $\theta$ and thus serves as a proxy for the R-symmetry of the IR theory.   The R-charges of the fields are then determined by c-extremization~\cite{Distler:1995mi,Melnikov:2009nh,Benini:2012cz,Melnikov:2016dnx} :   they $\GUL$ charges are normalized so that
\begin{align}
\label{eq:norm}
-\sum_{A} Q_A - \sum_i q_i  = \sum_A Q_A^2 - \sum_i q_i^2~,
\end{align}
and the R-charges of the multiplets are then
\begin{align}
\GUR:&&  \Gamma_A &\to e^{i(1+Q^A) \lambda} \Gamma_A  & \Phi_i &\to e^{i q_i\lambda} \Phi_i~.
\end{align}
We will assume that the UV R-symmetry current flows to the IR R-symmetry current, so that 
anomaly matching determines the central charge of the IR theory.  The result is
\begin{align}
\label{eq:cbar}
\frac{\cb}{3} = -\sum_{A} (1+Q_A) + \sum_i(1-q_i)~.
\end{align}
The gravitational anomaly fixes
\begin{align}
\label{eq:gravanom}
c- \cb = N-n~.
\end{align}

\subsubsection*{Constraints from the chiral algebra}

In addition to these basic characteristics, as in (2,2) case, in the absence of accidents we can also track the chiral algebra of the SCFT in terms of the UV data~\cite{Witten:1993jg,Kawai:1994qy,Melnikov:2009nh,Dedushenko:2015opz}.  The UV theory has a non-trivial cohomology for the supercharge $\bQb$, which acts as follows on the fundamental fields:
\begin{align}
\bQb \cdot \phi_i & = 0~, &
\bQb \cdot \phib_i & = \psib_i~,&
\bQb \cdot \psi_i & = -\pb \phi_i~,&
\bQb \cdot \psib_i & = 0~,\nonumber\\
\bQb\cdot \gamma_A & = 0~, &
\bQb\cdot\gammab_A & = -J_A(\phi)~.
\end{align}
The $\bQb$--cohomology has a well-understood subset:  the topological heterotic ring~\cite{Adams:2005tc}, which is described by a well-known algebraic construction:  the Koszul complex associated to the ideal $\la J_1,\ldots, J_N\ra \subset \C[\phi_1,\ldots,\phi_n]$~\cite{Kawai:1994qy,Melnikov:2009nh}.  The full $\bQb$--cohomology is more general and includes composite operators that involve the $\gamma_A$ fields.  For our purposes we will just glean three observations from that rich field:
\begin{enumerate}
\item the bosonic field $\phi_i$ is an element of the chiral algebra of R-charge $q_i$ unless some $J_A$ contains $\phi_i$ as a monomial;
\item the fermionic field $\gamma_A$ is an element of the chiral algebra of R-charge $1+Q_A$ unless it gains a mass due to a linear monomial just mentioned;
\item when all of the $\gamma_A$ are massless, since the operators $\gamma_A$ $\gamma_B$ have a non-singular OPE, we expect that the operator $\gamma_1 \gamma_2 \cdots \gamma_N$ is a well-defined element of the chiral algebra of R-charge $\sum_{A} (1+Q_A)$.
\end{enumerate}
Since we are interested in the IR limit of these LG theories, we will explicitly exclude mass terms from the superpotential.  Thus, if the flow from the LG theory to the IR is accident free, we expect that the SCFT will have chiral primary operators with the following charges and conformal weights:
\begin{align}
\xymatrix@R=2mm@C=2mm{
\text{operator}	& \GUL		&\GUR		& h				&\hb \\
\phi_i		& q_i			&q_i			& q_i/2 			& q_i/2 \\
\gamma_A	&Q_A		&1+Q_A		& (2+Q_A)/2		&(1+Q_A)/2 \\
\gamma_1\cdots \gamma_N
			&\textstyle\sum_{A} Q_A
			&\textstyle\sum_{A} (1+Q_A)
			&\textstyle\sum_{A} (2+Q_A)/2
			&\textstyle\sum_{A} (1+Q_A)/2
}
\end{align}
If these operators exist in the unitary IR SCFT, then the charges must satisfy the familiar unitarity bounds of the N=2 superconformal algebra:
\begin{align}
\label{eq:unibounds}
0 & < q_i \le \cb/3~,&
0 & < (1+Q_A) \le \cb/3~,&
0 & < \textstyle\sum_{A} (1+Q_A) \le \cb/3~.
\end{align}
The last constraint is particularly powerful.\footnote{These remarks have been made previously in~\cite{Bertolini:2014ela}.}  For instance, if we apply it to (2,2) theories, where $Q_i = q_i-1$, we find the relation
\begin{align}
\sum_i q_i \le \sum_i (1-2 q_i)~ \implies  n \le c~.
\end{align}
This bound is known to hold for all compact (2,2) LG theories~\cite{Kreuzer:1992bi}, but its direct derivation as a consequence of compactness is not straightforward.  Here we see that it follows if we assume that the SCFT realizes the chiral algebra.\footnote{As a further check of the consistency of these conditions, we studied the elliptic genus for compact (2,2) LG theories with $c=\cb=9$ and confirmed that each such theory has a unique operator with the quantum numbers of $\gamma_1\cdots\gamma_n$.}  As we will see in the next section, this bound will lead to powerful constraints on (0,2) LG theories as well.

\subsubsection*{Field redefinitions and equivalence relations}
Just as in the (2,2) theories, holomorphic field redefinitions compatible with the $\GUL$ charges lead to equivalence relations:  we assume that any two compact (0,2) LG theories related by such a field redefinition flow to the same fixed point.  In (0,2) theories there is a further complication that the redefinitions act separately on the $\Phi_i$ and on the $\Gamma_A$.  The former lead to changes of coordinates on the ideal, while the latter change the basis of generators of the ideal.  Furthermore, as in~\cite{Bertolini:2014ela}, any orbit of field redefinitions that contains a point with enhanced global symmetry can potentially lead to an accidental symmetry.  We will see examples of such accidental orbits, and we will take care to exclude them from our analysis.

\subsubsection*{Classification goal}
Having reviewed the setup, we can now precisely state our goal: we wish to classify compact (0,2) LG theories with $c=\cb <3$.  This means finding all quasi-homogeneous zero-dimensional ideals $\mathbb{J} = \la J_1,\ldots, J_n\ra \subset \C[\phi_1,\ldots,\phi_n]$ up to equivalence by holomorphic changes of variables and the basis of generators.  We will exclude accidental orbits and demand that the $\GUL$ charges satisfy~(\ref{eq:norm}) and~(\ref{eq:cbar})~.

\section{Ruling out theories with more than two fermi multiplets}\label{s:justtwo}
While the vanishing of the gravitational anomaly implies $n=N$, we do not yet have bounds on $n$.  In this section we will demonstrate that $n\le 2$ if $\cb <3$ and the LG theory yields an accident--free flow to a unitary theory.  This is a key point because the classification of zero-dimensional ideals $\mathbb{J}$ with $n$ generators is a difficult problem.  In the (2,2) case it is much simplified because the generators arise from the (2,2) superpotential $W$, so that the same combinatoric object---the Newton polytope for $W$---controls the generators. 

With our assumptions, the LG fields $\phi_i$ and $\gamma_A$ correspond to chiral primary operators in the SCFT.  When $\cb<3$ unitarity of the N=2 superconformal algebra implies that the charges of these fields satisfy
\begin{align}
\label{eq:smallunitarity}
\cb/3 & = \frac{K}{K+2}~,&
q_i    & = \frac{r_i}{K+2}~,&
1+Q_A & = \frac{P_A}{K+2}~,
\end{align}
where $K$, $r_i$, and $P_A$ are positive integers with $0<r_i \le K$ and $0<P_A\le K$.  

Since $-Q_A$ is the quasi-homogeneity charge of the polynomial $J_A$, it must be possible to find a  matrix of non-negative integers $M_A^i$ such that
\begin{align}
Q_A & = -\sum_{i} M_A^i q_i~,
\end{align}
and for any $i$
\begin{align}
\label{eq:Mconstraint}
\sum_{A} M_{A}^i \ge 2~.
\end{align}
If it is not possible to satisfy~(\ref{eq:Mconstraint}), then the $\mathbb{J}$ fails to satisfy our requirements:  if $\sum_{A} M_{A}^i =0$, then $\phi_i$ does not appear in the generators, and $\mathbb{J}$ cannot be zero-dimensional;  if $\sum_{A} M_{A}^i =1$, then $\phi_i$ appears linearly in precisely one generator, and that is a mass term. Of course the matrix $M_A^i$ is far from unique, but it must satisfy~(\ref{eq:Mconstraint}).

Given such an $M$, we see from~(\ref{eq:cbar}) that
\begin{align}
\sum_{A,i}  \left(M_{A}^i -\delta_{A}^i\right) r_i= K~,
\end{align}
and the last equation in~(\ref{eq:unibounds}) implies
\begin{align}
\sum_{A}\left(K+2 - \sum_i M_{A}^i r_i\right) \le K~.
\end{align}
Combining these two relations, we obtain
\begin{align}
(n-1) \sum_{A,i} (M_{A}^i-\delta_{A}^i) r_i - \sum_{A} (\sum_i M_A^i r_i -2) \le 0~.
\end{align}
Rearranging the terms, we obtain
\begin{align}
2n+ (n-2) \sum_i\left( \sum_{A} (M_A^i - \delta_A^i) \right) r_i - \sum_i r_i \le 0~.
\end{align}
Finally, if $n\ge 3$, then we obtain a necessary condition for this inequality to be true by using
~(\ref{eq:Mconstraint}) to replace $\sum_{A} (M_{A}^i -\delta_A^i)$ by its minimum value of $1$. 
This leads to
\begin{align}
2n + (n-3) \sum_i r_i \le 0~,
\end{align}
and we see that LG theories satisfying our assumptions with $n\ge 3$ cannot have an accident-free RG flow to an SCFT with $\cb<3$.

The $n=1$ case is quite familiar:  in this case $J_1 = \phi_1^{k-1}$ is equivalent by a field redefinition to the $\text{A}_{k-2}$ (2,2) LG theory.  What remains is a manageable classification problem of zero-dimensional quasi-homogeneous ideals $\la J_1, J_2\ra \subset \C[\phi_1,\phi_2]$ that also satisfy $\cb <3$.

\section{Constraints from compactness and charge bounds} \label{s:bounds}
As we remarked above, the potential of a (0,2) LG theory will be compact if and only if $\mathbb{J}$ is a zero-dimensional ideal.  An elementary result in commutative algebra~\cite{Cox:1998ua} is a necessary condition for zero-dimensionality:  monomials of the form $\phi_1^{k_1}$ and $\phi_2^{k_2}$ must appear in the generators.  Thus, a zero-dimensional ideal must belong to one of two classes:
\begin{align}
\underbrace{ J_1 \supset \phi_1^{k_1} + \phi_1^{l_1}\phi_2^{l_2}~, J_2 \supset \phi_2^{k_2}}_{\text{Class 1}}~~,&&
\underbrace{ J_1 \supset \phi_1^{k_1} + \phi_2^{k_2}~, J_2 \supset \phi_1^{l_1} \phi_2^{l_2}}_{\text{Class 2}}
~.
\end{align}
We will sometimes just specify the exponents for these two classes as $[k_1,k_2;l_1,l_2]_{1}$ and $[k_1,k_2;l_1,l_2]_1$.
More precisely, the indicated monomials must appear with generic non-zero coefficients in order to obtain a zero-dimensional ideal; of course other monomials may appear as well.  The mixed monomial also deserves a word of explanation:  if the $\phi_1^{k_1}$ and $\phi_2^{k_2}$ are the only ``pure'' monomials present in the ideal, then for Class 2  $J_2$ must contain the indicated mixed monomial---otherwise it will not be zero-dimensional.  For Class 1 the situation is slightly different:  if no mixed monomial can show up in either $J_1$ or $J_2$, then the theory is trivially equivalent to a sum of two (2,2) minimal models.   We will now apply the constraint $\cb<3$ to the two classes.

\subsection{Results of a numerical exploration}
To get our bearings, we simply scanned through a range of exponents $k_1,k_2$ and $l_1,l_2$ 
 found several families as well as exceptional cases with $\cb < 3$.  We summarize these in the following table.

\begin{align}
\label{eq:fulltable}
\xymatrix@R=1mm@C=40mm{
\text{Class 1} & \text{Class 2} \\
\text{a} : \la \phi_1^{k_1} + \phi_1\phi_2,\phi_2^{k_2}\ra
&
\text{a} : \la \phi_1^{k_1} + \phi_2^{k_2},\phi_1\phi_2\ra  \\
\text{a}^\ast : \la \phi_1^{k_1} + \phi_1^{k_1-1}\phi_2^{k_2-1},\phi_2^{k_2}\ra
&
\text{a}': \la \phi_1^2 + \phi_2^2, \phi_1^{l_1} \phi_2^{l_2}\ra \\
\text{b}: \la \phi_1^{k_1} + \phi_1^2\phi_2,\phi_2^2\ra 
&
\text{b}:  \la \phi_1^{k_1}+\phi_2^{2},\phi_1^2\phi_2\ra~ \\
\text{b}^\ast: \la \phi_1^2+\phi_1\phi_2^{k_2-2},\phi_2^{k_2}\ra
&
~ \\
\text{c}: \la\phi_1^3+\phi_2^2,\phi_2^3 \ra  
&
\text{c}: \la\phi_1^3 + \phi_2^2,\phi_1^3\phi_2\ra \\
\text{d}: \la \phi_1^2 +\phi_2^3,\phi_2^4 \ra &
\text{d} : \la \phi_1^3+\phi_2^2,\phi_1^2\phi_2^2\ra \\
\text{e} : \la \phi_1^2+\phi_2^3,\phi_2^5\ra &
\text{e}: \la \phi_1^3+\phi_2^2,\phi_1\phi_2^2\ra \\
}
\end{align}
Among these classes we recognize some familiar theories:  the $\GE_7$ minimal model is equivalent up to field redefinitions to Class 2e, while the $\text{D}_{k-1}$ minimal models are a subset of Class 2a with $k_2 = 2$.  In the next section we will show that every Class 1 or 2 theory belongs to this list.  We will then explore equivalences among these different cases and produce a substantially shorter list.

\subsection{Class 1 and $\cb<3$}
The first step in analyzing this case is to establish some a priori upper bounds on the exponents.  First, $k_1 > l_1$: otherwise $\phi_2$ will have a negative R-charge.   Second, without loss of generality $l_2 < k_2$: otherwise applying a field redefinition 
\begin{align}
\Gamma_2 \to \Gamma_2 - \phi_1^{l_1} \phi_2^{l_2-k_2} \Gamma_1
\end{align}
decreases the power of $\phi_2$ of the mixed monomial in $J_1$.\footnote{Either $J_2$ has no additional monomials, or it has a mixed monomial with a smaller power of $\phi_2$.  The first possibility reduces the ideal to the familiar product of (2,2) models, while the second possibility allows us to decrease the $l_2$ exponent in the mixed monomial in $J_1$.  We can apply the field redefinition until the desired $l_2 < k_2$ bound is reached.}  Of course we also have $l_2>0$.

With those bounds in place, we turn to the central charge itself.  A short computation leads to
\begin{align}
\frac{\cb}{3} = \frac{ \left( l_2(k_1-1) + (k_1-l_1) (k_2-1) \right)^2}{l_2^2(k_1^2-1)+(k_1-l_1)^2(k_2^2-1)}~.
\end{align}
It is convenient to redefine the integers:
\begin{align}
k_2 & = s_2+1~,&
k_1 & = l_1 + m_1~.
\end{align}
The condition $\cb < 3$ is then equivalent to the positivity of the following function of integers
\begin{align}
F = l_2^2(l_1+m_1-1) + \left(m_1^2-l_2 m_1(l_1+m_1-1)\right) s_2 > 0~
\end{align}
on the domain defined by the inequalities
\begin{align}
\label{eq:domaindefine}
0 & \le l_1~,&
s_2 &\ge l_2 > \begin{cases} 1 &\text{if} ~~l_1 = 0 \\ 0 &\text{if}~~l_1>0 \end{cases},& 
m_1 &> \begin{cases}  1 & \text{if} ~~ l_1 = 0 \\ 0 & \text{if}~~ l_1 >0 \end{cases}~.
\end{align}
The special cases for $l_1=0$ avoid mass terms.

We find it convenient to examine $l_1 = 0$ separately from the remaining possibilities.  Here
\begin{align}
F = l_2^2(m_1-1) -m_1 Z s_2~,
\end{align}
where
\begin{align}
Z =  l_2(m_1-1) -m_1~.
\end{align}
On the domain defined by the inequalities~(\ref{eq:domaindefine}) $Z$ is non-negative, and $Z=0$ only if $l_2 =2$ and $ m_1 = 2$.  When $Z=0$ $F>0$ for all values of $k_2$, and we encounter our first infinite family of $\cb < 3$ (0,2) LG theories:
\begin{align}
[2,k_2;0,2]_1  \longleftrightarrow \mathbb{J} = \la \phi_1^2+\phi_2^2, \phi_2^{k_2}\ra~.
\end{align}
This belongs to Class $\text{2a}'$.

Next, we assume $Z>0$, so that $F>0$ is equivalent to 
\begin{align}
\frac{s_2}{l_2} < \frac{l_2(m_1-1)}{m_1(l_2 (m_1-1)-m_1)} = \frac{1}{m_1} + \frac{ 1}{l_2(m_1-1)-m_1}~.
\end{align}
Since $s_2 \ge l_2$, the right-hand-side must be greater than $1$, and the only solutions are 
\begin{align}
(m_1,l_2)\in\{(2,3), (3,2)\}~.
\end{align}
Plugging these choices back into $F$, we find that $s_2$ is bounded, and there are just three integer solutions:  1c, 1d, and 1e.

Next, we consider $l_1>0$, where 
\begin{align}
F& =  l_2^2(l_1+m_1-1) + Y s_2 ~, &  Y & = m_1\left(m_1-l_2 (l_1+m_1-1)\right)~.
\end{align}
If $Y\ge 0$ then $F$ is positive.  Since $m_1>0$ and $l_1+m_1>1$, this requires $l_2 \le 1$.  Thus, $Y\ge 0$ precisely when $l_1 = l_2 = 1$.  This is the family 1a in the list above.

When $l_2>1$, then $Y<0$, and thus $F>0$ again puts a bound on the ratio $s_2/l_2$, which leads to the constraint
\begin{align}
1 <\frac{1}{m_1} + \frac{1}{l_2(l_1+m_1-1)-m_1}~.
\end{align}
This can only be achieved if $m_1 =1$ or if $l_2(l_1+m_1-1)-m_1 =1$.  The former possibility leads to classes $\text{1a}^\ast$ and $\text{1b}^\ast$, and the latter possibility leads to class $\text{1b}$.

\subsection{Class 2 and $\cb <3$}
We turn to Class 2 theories and verify that every set of exponents consistent with $\cb <3$ belongs to one of the theories in~(\ref{eq:fulltable}).   There is a symmetry of the Class 2 theories that exchanges $\phi_1$ and $\phi_2$, as well as the corresponding exponents; we can therefore focus on theories with $k_1\ge k_2$.

For Class 2 the central charge is given by
\begin{align}
\frac{\cb}{3} & = \frac{ (k_1 k_2 + k_2(l_1-1)+k_1(l_1-1))^2}{k_1^2 k_2^2-k_1^2-k_2^2+(k_1l_2+k_2l_1)^2}~.
\end{align}
Imposing $\cb < 3$ is then equivalent to the positivity of
\begin{align}
F = k_1k_2 - (k_1 k_2 -k_1-k_2) \left[ k_1(l_2-1)+k_2(l_1-1)\right]~.
\end{align}
If $l_1 = l_2 = 1$ or  $k_1 = k_2 = 2$ then $F>0$.  We recognize these $\cb<3$ solutions as Class 2a and $\text{2a}'$, respectively.  

To study the remaining possibilities, we can assume that $k_1\ge 3$ and $k_2 \ge 2$, and that either $l_1$ or $l_2$ is greater than one.  With those assumptions it is evident that $(k_1k_2-k_1-k_2)$ is positive, and 
\begin{align}
k_1(l_2-1)+k_2(l_1-1) \ge 5~.
\end{align}
Thus, 
\begin{align}
F \le k_1k_2 -5(k_1k_2-k_1-k_2) = 5(k_1+k_2) - 4 k_1 k_2 \implies \frac{k_1}{k_2} \le \frac{5}{4 k_2-5}~.
\end{align}
But, since we also assume $k_1 \ge k_2$, this immediately implies $k_2 = 2$.  Setting $k_2=2$, we find
\begin{align}
F = 2 k_1 - (k_1-2)\left[ k_1(l_2-1)+2(l_1-1)\right]~,
\end{align}
and using $k_1 \ge 3$ in the second term, $F>0$ implies 
\begin{align}
k_1(3 -l_2) - 2(l_1-1) >0~.
\end{align}
Thus $l_2 <3$.  Setting $l_2 = 1$, we now find
\begin{align}
F = 2 \left[ k_1 - (k_1-1)(l_1-1) \right]~.
\end{align}
Thus, we obtain $\cb < 3$ theories for: $[k_1,2;1,1]_2$---these are a special case of class 2a theories; $[k_1,2;2,1]_2$--these are the class 2b theories; $[3,2;3,1]_2$---the class 2c theory.   

The last case to consider is $l_2 = 2$ and $k_2 = 2$, for which
\begin{align}
F = 2 k_1 - (k_1-2)(k_1-2 + 2 l_1)~.
\end{align}
Since $k_1\ge 3$ and $l_1 \ge 1$, we see that
\begin{align}
F & < k_1 - 2(l_1-1)~ &\quad\text{and}\quad
F & <6-k_1~.
\end{align}
This is a finite list of possibilities to check, and the $\cb <3$ exponents are $[3,2;1,2]_2$ and $[3,2;2,2]_2$---the 2e and 2d classes, respectively.

This completes the proof that the exponents in~(\ref{eq:fulltable}) include every $\cb <3$ theory.  In the next section we will show that this list is redundant:  many of the classes are related to each other by field redefinitions.

\section{Redefinitions and classification results} \label{s:classredef}
The general form of the (0,2) superpotential is determined once the charges $Q_A$ and $q_i$ are fixed.  For each $A$ the monomials  $\prod_i \phi_i^{m_i}$  in $J_A$ belong to a Newton polytope $\Delta_A$ that lies in the positive orthant of $\R^n$.  The intersection of $\Delta_A$ with the underlying lattice $\Z^n \subset \R^n$ determines the exponents of the monomials in $J_A$.  Quasi-homogeneity implies that the $\Delta_A$ lie in parallel hyperplanes in $\R^n$.  With this combinatorial set-up, we can write the generic superpotential as
\begin{align}
J_A = \sum_{m \in \Delta_A \cap \Z^n} \alpha_m \prod_{i=1}^n \Phi_i^{m_i}~.
\end{align}
The coefficients $\alpha_m$ determine a point in   $\cM_0 = \C^{\sum_A |\Delta_A|}$, where $|\Delta_A|$ denotes the number of points in $\Delta_A \cap \Z^n$.

Compactness implies that for a generic point $p\in \cM_0$, the ideal is zero-dimensional, but there will be a locus $\cZ\subset \cM_0$ where the theory is non-compact.   This singular locus is preserved by field redefinitions, so that we can form
\begin{align}
\cM_{\text{uv}} = \frac{\cM_0 \setminus \cZ}{\text{field redefinitions}}~.
\end{align}
On a first look we can think of $\cM_{\text{uv}}$ as a model for the moduli space of the IR SCFT, but there are a number of issues with this simplistic view.
\begin{enumerate}
\item $\cM_{\text{uv}}$ is complicated because the quotient is in general not reductive.  Even for a reductive quotient, one has to provide stability conditions to obtain a well-behaved (though perhaps singular) space.  Similar issues arise already in a simpler toric setting~\cite{Bertolini:2014ela,Donagi:2014koa}.
\item The space $\cM_{\text{uv}}$ in general includes points with enhanced continuous global symmetries, and these can invalidate the identification of the R-symmetry~\cite{Bertolini:2014ela} and therefore the correspondence between the chiral UV and IR data.  
\item Some of the deformations may be identified in the IR, even though they correspond to distinct motions in $\cM_{\text{uv}}$.
\item It may be that the IR theory has additional marginal deformations that cannot be realized by deformations of the (0,2) superpotential.  This and the previous complication are already familiar in the context of (2,2) linear sigma models that flow to compact (2,2) SCFTs---see, for example,~\cite{Cox:2000vi,Kreuzer:2010ph}.
\end{enumerate}
This general discussion simplifies a great deal when applied to the (0,2) LG theories studied here.  On general grounds a compact unitary $\cb< 3$ SCFT cannot have marginal supersymmetric deformations:  the corresponding operator must be chiral primary with charge $\qb = 1$, but such operators are forbidden by the unitarity bound $\qb \le \cb/3$.  The other simplification is that the enhanced continuous symmetry locus in $\cM_0\setminus\cZ$ is easy to describe:  it consists of all points that are equivalent to the ideal
\begin{align}
\mathbb{J}_{\text{ac}} = \la \phi_1^{k_1},\phi_2^{k_2}\ra~.
\end{align}
All theories on this locus will be equivalent to a product of two (2,2) $A_{k}$ minimal models.

In view of the preceding discussion, to complete our classification, we will make three steps.  First, we will identify families of theories in~(\ref{eq:fulltable}) with unique charges for the chiral fields.  Second, for each family we will show that unless $p \in \cM_{\text{uv}}$ belongs to an enhanced symmetry orbit, we can use field redefinitions to map the entire family to a single point:  in other words the deformation space is trivial, as it should be for a $\cb < 3$ theory.  Finally, we will verify that each such theory satisfies the unitarity bounds~(\ref{eq:smallunitarity}).

\subsection*{Isomorphic families}
To start, we note the isomorphic families $\text{1a} = \text{1a}^\ast$, $\text{1b} = \text{1b}^\ast$, $\text{1c} = \text{2c}$, $\text{1d} = \text{2d}$, and $\text{1e} = \text{2e}$. This follows by simply comparing the charges.  For instance, to demonstrate $\text{1a} = \text{1a}^\ast$ we compare
\begin{align}
\text{1a} :&&
q_1 &= \frac{k_2}{k_2^2(k_1-1)+2}~,&
q_2 &= (k_1-1) q_1~, &
Q_1 & = - k_1 q_1~, &
Q_2 & = -k_2(k_1-1)q_1
\end{align}
with
\begin{align}
\text{1a}^\ast :&&
q_2 &= \frac{k_1}{k_1^2(k_2-1)+2}~,&
q_1 &= (k_2-1) q_2~, &
Q_2 & = - k_2 q_2~, &
Q_1 & = -k_1(k_2-1)q_2~.
\end{align}
Evidently, the two sets of charges are the same up to exchanging $\phi_1$ with $\phi_2$ and $J_1$ with $J_2$.  Similar comparisons of the charges lead to the other isomorphisms.

In fact, there are just $5$ distinct families to consider, and we can pick representatives exclusively from Class 2.  This is because $\text{2a'} \subset \text{2a}$, $\text{1a} \subset \text{2a}$, and $\text{1b} \subset \text{2b}$.  More precisely, we find isomorphic charges for
\begin{align}
[l_1+l_2, l_1+l_2; 1,1]_2 \simeq [2,2;l_1,l_2]_2~,&&
[k_2(k_1-1),k_2;1,1]_2 \simeq [k_1,k_2;1,1]_1~, \nonumber\\
[2k_1-4,2;2,1]_2 \simeq [ k_1,2;2,1]_1~.
\end{align}
It is also not too hard to see that $[3,3;1,1]_{2} \simeq [2,2;2,1]_2$, which means we can take the integer in class 2b to satisfy $k_1\ge 3$.

\subsection*{Trivial parameter spaces}
We have now shown that every (0,2) LG theory with $c=\cb <3$ and without an accidental symmetry belongs to one of 5 classes:  2a, 2b, 2c, 2d, or 2e.  The first of these depends on two integer parameters, which we may as well take to satisfy $k_1\ge k_2> 2$;\footnote{Recall that $k_2=2$ just corresponds to the (2,2) D-series.} the second depends on one integer parameter $k_1 \ge 3$.

Each of these classes has a trivial parameter space in the sense described above:  every deformation of the superpotential can be undone by a field redefinition.  We will now show this in some detail for class 2a and give a sketch for class 2b; the remaining classes 2c, 2d, and 2e are simpler since they do not involve integer parameters, and we leave the details to the interested reader.

\subsubsection*{Class 2a}
In class 2a the ideal has the form
\begin{align}
J_1 &\supset \phi_1^{k_1} + \phi_2^{k_2}~,&
J_2 & \supset \phi_1 \phi_2~.
\end{align} 
Suppose $J_2$ admits a monomial $\phi_1^{m_1} \phi_2^{m_2}$.  Since the R-charges of $\phi_1$ and $\phi_2$ must be positive, this monomial cannot be divisible by $\phi_1\phi_2$.  Thus, $J_2$ must be of the form
\begin{align}
J_2 = \alpha_1 \phi_1\phi_2 + \alpha_2 \phi_1^{m_1}+ \alpha_3 \phi_2^{m_2} ~.
\end{align}
By examining the charges of the monomials, we see that all three monomials cannot be present in $J_2$ simultaneously if $k_2 >2$.  Thus, we must have either $\alpha_3  = 0$ and $q_2 = q_1(m_1-1)$, or $\alpha_2=0$ and $q_1 = q_2(m_2-1)$.  Since $q_1 k_1 = q_2 k_2$, it must be that $k_1 = k_2(m_1-1)$ in the first case, and $k_2 = k_1(m_2-1)$ in the second case.  We need not consider the last possibility because of our a priori assumption $k_1 \ge k_2$.

It may be that $k_2$ does not divide $k_1$.  In this case $J_2 = \alpha_1 \phi_1\phi_2$, and $\alpha_1 \neq 0$ for compact theories.  This means that all mixed monomials allowed by the charges in $J_1$ can be removed by a field redefinition of the form
\begin{align}
\Gamma_2 \to \Gamma_2 - P(\phi_1,\phi_2) \Gamma_1~,
\end{align}
where we choose $P$ such that $\phi_1\phi_2 P$ is the sum of the mixed monomials in $J_1$.  So, in this situation the most general zero-dimensional ideal is equivalent to
\begin{align}
\label{eq:boring2a}
J_1 &= \beta_1 \phi_1^{k_1} + \beta_2\phi_2^{k_2}~,&
J_2 & = \alpha_1 \phi_1\phi_2~,
\end{align}
with non-zero coefficients $\alpha_1$, $\beta_1$, and $\beta_2$.  These can be absorbed by rescaling $\phi_1$, $\phi_2$, and $\Gamma_2$.

Next, we set $k_1 = p k_2$ for some positive integer $p$.   Now the most general ideal is of the form
\begin{align}
\label{eq:class2adivisible}
J_1 & = \beta_1 \phi_1^{k_1} + \beta_2 \phi_1^{k_1-1} \phi_2^p + \cdots, \nonumber\\
J_2 & =  (\alpha_1 \phi_1 + \alpha_2 \phi_2^{p})\phi_2~.
\end{align}
Clearly $\beta_1 \neq 0$ if the ideal is zero-dimensional.  If also $\alpha_1\neq 0$, then we can redefine $\phi_1\to \phi_1+\xi \phi_2^p$, to obtain an equivalent ideal of the form
\begin{align}
J_1 & = \beta_1 \phi_1^{k_1} + \beta_2 \phi_1^{k_1-1} \phi_2^p + \cdots, \nonumber\\
J_2 & =  \alpha_1 \phi_1 \phi_2~.
\end{align}
By the same argument as in the previous paragraph, this case is then equivalent to~(\ref{eq:boring2a}).  On the other hand, if $\alpha_1 = 0$, we have an ideal
\begin{align}
J_1 & = \beta_1 \phi_1^{k_1} + \beta_2 \phi_1^{k_1-1} \phi_2^p + \cdots, \nonumber\\
J_2 & =   \alpha_2 \phi_2^{p+1}~,
\end{align}
with $\beta_1\neq0$ and $\alpha_2\neq0$.  By using a combination of $\phi_1$ and $\Gamma_2$ redefinitions, we then find this ideal is equivalent to $J_1 = \phi_1^{k_1}$ and $J_2 = \phi_2^{p+1}$.   This is clearly an enhanced symmetry orbit, and we expect that these theories will simply flow to a product of two $(2,2)$ minimal models.


\subsubsection*{Class 2b}
Now we perform an analogous analysis for class 2b, for which the ideal is of the form
\begin{align}
J_1 &\supset \phi_1^{k_1} + \phi_2^2~, &
J_2 &\supset \phi_1^2\phi_2~,
\end{align}
and $k_1 \ge 3$.  If $k_1 = 3$, there are no additional monomials in either generator, and correspondingly, there are no parameters except trivial coefficients that can be absorbed into rescaling the fields.  The next possibility, $k_1 = 4$, allows for two additional monomials in $J_2$; indeed, in this case $J_1$ and $J_2$ can be taken to be generic polynomials of degree $2$ in $\phi_2$ and $\phi_1^2$.  Once again, any zero-dimensional ideal is either on an enhanced symmetry orbit or is equivalent to $\la \phi_1^4 + \phi_2^2,\phi_1^2\phi_2\ra$.  This is perhaps not entirely obvious, so let us elaborate on it.

The generic ideal takes the form
\begin{align}
J_1 & = \beta_1 \phi_2^2  + \beta_2\phi_1^4+ \beta_3 \phi_1^2 \phi_2~, \nonumber\\
J_2 & = \alpha_1 \phi_2^2  + \alpha_2\phi_1^4+ \alpha_3 \phi_1^2 \phi_2~.
\end{align}
A non-singular ideal is equivalent to the ideal with $\alpha_1 = 0$ and $\beta_1 = 1$.  If, in addition, $\alpha_3 = 0$, then the ideal belongs to an enhanced symmetry orbit.  On the other hand, if $\alpha_3 \neq 0$, then, by using a redefinition $\phi_2 \to \phi_2 +\gamma\phi_1^2$, followed by a rescaling of $\Gamma_2$, we see that  the ideal is equivalent to
\begin{align}
J_1 &= \phi_2^2 + \beta_2 \phi_1^2 \phi_2 + \beta_3\phi_1^4~,\nonumber\\
J_2 &= \phi_1^2\phi_2~.
\end{align}
Finally, the redefinition $\Gamma_2\to \Gamma_2 -\beta_2 \Gamma_1$ eliminates the mixed monomial in $J_1$, and the result is indeed equivalent to $\la \phi_1^4 + \phi_2^2,\phi_1^2\phi_2\ra$.

For $k_1>4$ the only possible additional monomial in $J_2$ is $\phi_1^{m_1}$, and it is allowed only if $k_1 = 2(m_1-2)$.  It is also the case that the generator $J_1$ has an additional monomial only when $k_1$ is even.  So, setting $k_1 = 2p$ with $p\ge 3$, we have the general form of the generators as 
\begin{align}
J_1 &= \beta_1 \phi_1^{2p} + \beta_2 \phi_1^{p}\phi_2 +\beta_3\phi_2^2~,&
J_2 & =\phi_1^2( \alpha_1 \phi_2 +\alpha_2 \phi_1^p)~.
\end{align}
By an analogous argument to that given after~(\ref{eq:class2adivisible}) for class 2a, it now follows that every zero-dimensional ideal that is not on an enhanced symmetry orbit is equivalent to $\la \phi_1^{k_1} +\phi_2^2,\phi_1^2\phi_2\ra$.

\section{Discussion}
We classified compact (0,2) LG theories that flow to IR fixed points with $c=\cb <3$.  Such a theory must either be one of the  familiar (2,2) ADE theories, or it must belong to one of our classes.  In the latter case the holomorphic data is specified by a zero-dimensional quasi-homogeneous ideal $\mathbb{J} \subset \C[\phi_1,\phi_2]$, and up to field redefinitions, these ideals belong to one of the following classes
\begin{align}
\label{eq:thefinalcut}
\xymatrix@R=1mm@C=2mm{
\text{ideal} & q_1 & q_2 & c/3 \\
\la \phi_1^{k_1} + \phi_2^{k_2}, \phi_1\phi_2\ra~,~~ k_1\ge k_2 > 2~&
\frac{k_2}{k_1 k_2+2} &
 \frac{k_1}{k_1 k_2 +2} &
\frac{k_1k_2}{k_1k_2+2} \\ 
\la \phi_1^{k} + \phi_2^2, \phi_1^2\phi_2 \ra~,~~ k\ge 3\qquad &
\frac{k+1}{(k+1)^2+2} &
\frac{k(k+1)}{2( (k+1)^2+2)} &
\frac{ (k+1)^2}{(k+1)^2+2} \\
 \la\phi_1^3 + \phi_2^2,\phi_1^3\phi_2\ra  &
 \frac{10}{52} & \frac{15}{52} & \frac{50}{52} \\
 \la \phi_1^3+\phi_2^2,\phi_1^2\phi_2^2\ra &
 \frac{22}{123} & \frac{33}{123} & \frac{121}{123}
}
\end{align}
The first of these classes was studied in some length in~\cite{Gadde:2016khg}, where persuasive arguments were given for identifying the IR theory with a specific chiral $\SU(2)/\GU(1)$ WZW coset SCFT.   It should be possible to extend those results to the remaining theories described in this note and explicitly identify the putative IR SCFTs.  This would offer additional checks to ensure that these theories describe accident--free RG flows and further enlarge the class of ``(0,2) minimal models''~\cite{Gadde:2016khg,Berglund:1995dv} that are, on one hand solvable, and on the other hand have simple UV realizations.

Even without making such detailed comparisons with putative IR realizations, we can see that these theories satisfy some non-trivial consistency checks.  For instance, we observed that, as required for $\cb<3$ theories, the (0,2) LG superpotentials for these theories are parameter free.  Another consistency check is that the charges of chiral operators satisfy the unitarity bounds~(\ref{eq:smallunitarity}) in a non-trivial way:  table~(\ref{eq:fulltable}) was obtained by just requiring ideals that are zero dimensional and satisfy $\cb<3$, $q_{1,2} >0$; however, all of the charges obtained satisfy the more stringent bounds of~(\ref{eq:smallunitarity}).

Our results have an interesting application to the construction of more general (0,2) theories at higher central charge, and it would be relatively straightforward to perform a taxonomic exercise of the sort familiar from~\cite{Lutken:1988hc} to describe the models that lead to $c=\cb = 9$ or $c=\cb = 6$; a projection onto integral R-charges can then produce potentially new (0,2) heterotic compactifications that are amenable to analysis following standard LG techniques~\cite{Distler:1993mk}.

One of the main technical simplifications in our analysis is the restriction to $n\le 2$, which we showed is a consequence of unitarity.  It would be interesting to explore the implications of this further.  In particular, is it possible to choose a zero-dimensional with $n\ge 3$ for which the naive application of extremization leads to $\cb <3 $?  What is the low energy limit of such a theory?  Is every such ideal on an enhanced symmetry orbit and therefore suffers from an accident, or is there an accident that cannot be detected by considering orbits of field redefinitions? 

It would also be useful to make more precise the argument given for the existence of the operator $\gamma_1\cdots \gamma_n$ in $\bQb$-cohomology.  Following~\cite{Witten:1993jg,Dedushenko:2015opz}, we expect that the $J_A$ dependence of the singularities in the OPE of $\bQb$--closed operators is $\bQb$-exact: we can compute the OPE of operators in $\bQb$-cohomology in the free theory, or, equivalently, in the $bc$--$\beta\gamma$ realization~\cite{Fre:1992hp}.   It would be useful to explore this point further:  for instance, what are the operators of the form $\gamma_1\cdots\gamma_n f(\phi)$ that belong to the $\bQb$-cohomology?  
These questions can be investigated by following the methodology laid out in~\cite{Dedushenko:2015opz}.

Another direction in which this work could be extended is to consider $c>\cb$ and $\cb <3$ theories.  It would be interesting to classify such LG theories.  Is there, for instance, a bound on $c -\cb$ above which there are no accident--free (0,2) LG flows?  We know from~\cite{Gadde:2016khg} that there are families of theories with $c>\cb$ and $\cb<3$, but we have no sense of how large the class of such theories might be.  This would be a fruitful direction for investigation since it would a richer holomorphic sector while still having the constraints of unitarity from $\cb <3$.  Such efforts may well yield new insights into (0,2) LG and (0,2) SCFT.

\bibliographystyle{./utphys}
\bibliography{../../BIB/newref}

\end{document}